# Pair Programming Practiced in Hybrid Work

Anastasiia Tkalich
*SINTEF*
Trondheim, Norway
anastasiia.tkalich@sintef.no

Nils Brede Moe
*SINTEF*
Trondheim, Norway
nils.b.moe@sintef.no

Nina Haugland Andersen
*Norwegian University of Science and Technology (NTNU)*
Trondheim, Norway
nina.haugland@ntnu.no

Viktoria Stray
*University of Oslo, SINTEF*
Oslo, Norway
stray@ifi.uio.no

Astri Moksnes Barbala
*SINTEF*
Trondheim, Norway
astri.barbala@sintef.no

*Abstract*—Pair programming (PP) has been a widespread practice for decades and is known for facilitating knowledge exchange and improving the quality of software. Many agilists advocated the importance of collocation, face-to-face interaction, and physical artifacts incorporated in the shared workspace when pairing. After a long period of forced work-from-home, many knowledge workers prefer to work remotely two or three days per week, which is affecting practices such as PP. In this revelatory single-case study, we aimed to understand how PP is practiced during hybrid work when team members alternate between on-site days and working from home. We collected qualitative and quantitative data through 11 semi-structured interviews, observations, feedback sessions, and self-reported surveys. The interviewees were members of an agile software development team in a Norwegian fintech company. The results presented in this paper indicate that PP can be practiced through on-site, remote, and mixed sessions, where the mixed mode seems to be the least advantageous. The findings highlight the importance of adapting the work environment to suit individual work mode preferences when it comes to PP. In the future, we will build on these findings to explore PP in other teams and organizations practicing hybrid work.

## I. Introduction

Since the Covid-19 outbreak, several studies have investigated the impact of work-from-home and other work modes on software development. For example, one extensive study of software engineers during the pandemic indicated that remote work could negatively impact productivity [1]. Although subsequent research has painted a more nuanced picture of this connection [2], it is evident that the work mode affects interaction within software teams [3] and their work practices. At the same time, we can assume that future software development will be hybrid, meaning that team members will be more likely to collaborate across different work modes [4].

Pair programming (PP) is one of the practices that is affected by the work mode [5]. PP is one of the core practices in agile methodologies. It is appreciated by developers and has been shown to improve satisfaction and productivity [6], [7], helps establish backup behavior in teams [8] and improves communication [9]. In this study, we explore the role of PP in modern software development, with particular attention to its implementation in hybrid work. More specifically, we seek to answer the following research question:

*How is pair programming practiced in hybrid work?*

By addressing this question, our goal is to provide valuable insights and practical recommendations for employers navigating hybrid software development. Furthermore, as the majority of PP research focuses on student-based studies (e.g., [10]–[16]), we aim to contribute to the field by providing an industrial case study.

## II. Background

### A. Pair programming

PP is a key agile practice that is believed to improve team performance as it supports effective teamwork through monitoring, feedback, and backup behavior [17]. In a typical PP session, two developers sit side-by-side at one computer, continuously collaborating on the same design, algorithm, code, or test [7]. One developer takes a leading role (called the driver) while another (called the navigator) observes and actively provides feedback, asks questions, and makes suggestions to ensure the high quality of the produced code [18]. Although the traditional roles of navigator and driver have been widely accepted, today, pairs often take on both responsibilities at the same time instead of having an explicit division of labor. Chong and Hurlbutt [19] found in their ethnographic study that having a strict separation of roles inhibits the natural way of working and that both developers having access to the keyboard enabled rapid switching and made the developers more engaged. Further, Wray [20] reports additional scenarios of how two developers can collaborate on jointly improving the same code, for example, by reviewing and discussing issues and potential solutions without any explicit roles. During a PP session, the majority of interaction happens through explaining and decision-making [21].

Several studies have compared the effectiveness of PP versus solo programming. In a meta-analysis, Hannay et al. [22] found that PP scored slightly higher both on the duration of work sequences and the quality of complex tasks than solo-programing. However, the latter also meant that PP required more effort than solo programming. Additionally, their analysis revealed that PP could have a time gain on simpler coding tasks, yet with lower quality as a cost. Further, they conclude that the effectiveness of PP depends on complex factors such as task complexity, motivation, and team climate. PP contributes to knowledge transfer and requires skills beyond software development skills, such as communication skills, as reported by Zieris and Prechelt [21]. The benefits of PP appear to go beyond a single programming task, as the knowledge transfer contributes to long-term effects among the practitioners (such as increased competition and mistakes avoided in the future). However, it is important to ensure a

The work was partially supported by the Research Council of Norway through the 10xTeams project (Grant 309344).

fluent PP process and complementary knowledge of the coding partners [23].

### B. Hybrid software development

Hybrid work has become a phenomenon of interest during and especially after the Covid-19 outbreak. On-site was no longer the default mode, while remote work did not fully establish [4]. Recent surveys show that employees prefer to work 2-3 days per week from home [24]. Hybrid work can be generally defined as a spectrum of flexible work arrangements in which some employees work mostly or completely remotely, others mostly or completely from the traditional office, and others in some combination of the two [4]. A well-known example of hybrid work is partially dispersed teams. Earlier research suggests that members working remotely in such teams experience reduced team cohesions, a poor overview of the team tasks, team coordination problems, and even conflicts [25]. In hybrid work, employees still appear in the office, but because many alternate office days with remote days, they do not necessarily meet. In other words, the members' office presence is not fully aligned [26]. Teams that are not aligned may experience lower psychological safety [3] and thus have a risk of decreased well-being and performance. Finally, meetings during hybrid work vary in terms of work mode and can be either *remote* (all participants are off-site), *on-site* (all participants are on-site), or *mixed* (some participants are on-site while others are remote) [27].

### C. Pair programming in different work modes

A lot of knowledge on PP comes from studying pairs working on-site, however remote PP has also been explored. Remote or distributed pair programming (DPP) entails two developers working remotely on the same design, algorithm, or code [28]. Research studying this mode of PP concludes that it has similar advantages as PP on-site, such as increased self-confidence and communication skills [13], [14]. However, a prerequisite for successful DPP is that the partners are familiar with each other before the distributed session(s) and that they have fairly good communication skills from the start in order to avoid misunderstandings [29]. The remote nature of collaboration can also contribute to challenges related to coordination difficulties (e.g., due to a mismatch in time schedules of the partners) [30], and technical hurdles (such as bad internet connection, websites crashing, and poor technical equipment at home compared to that at the office) [16]. Remote PP has also been explored in the context of work-from-home (WFH) during the COVID outbreak [5]. The authors found that the engineers who used to PP in the office stopped using the practice when they were forced to WFH. The authors suggest that remote collaboration is not as natural as that in the office and that the success of remote collaboration highly relies on social connections prior to the outbreak. Our study adds to the existing literature by examining how PP is practiced not only in the fully remote mode but also in the settings of hybrid work.

## III. RESEARCH DESIGN

To answer our research question, we designed a revelatory single-case study [31]. This approach is recommended when the phenomenon under investigation is relatively little known and thus needs to be approached exploratively. In our case, the phenomenon was PP in the hybrid work.

### A. Investigated company and team

*Company*: The study was carried out within SpareBank 1 Utvikling (SB1U), which is a Norwegian software company owned by an alliance of banks that employs 24 software teams. This company was interesting for our study because it is a mature agile organization that had been for many years focusing on developing health work practices, including PP, as reported in one of our previous studies [32]. SB1U is considered one of the most innovative technology companies in Norway, with the top-rated mobile bank app in Apple's App Store. SB1U had worked for years on scaling the software development capacity internally by building an attractive workplace. SB1U moved away from its legacy monolithic technical architecture several years ago, which was typical for banks, towards a microservice architecture. A modular architecture, tools, and automation enabled teams to have end-to-end responsibility and decision-making authority for their products, avoid handovers between teams and be able to develop software continuously.

*Team*: The team can be described as a platform team and was mostly responsible for the operations and maintenance of the applications used by the rest of the teams at SB1U. At the time of data collection, the team's priority was a migration of the applications to the cloud (AWS). The team relied on objectives and key results to guide their work and "Monday commitments" (MC) and "Friday wins" (FW) as key team meetings. During the MCs, the team planned their tasks and discussed the responsibilities for the week. The FWs were about celebrating what the team had achieved and about sharing learning. Additionally, the team had a meeting on Wednesdays where the agenda varied according to the current priorities. The team consisted of 15 members with 14 members alternating between days on-site and remote and one person working strictly offsite (see more details in the Results). In terms of experience in the bank, some members had worked in the company for less than two years (N=8), some between two and five years (N=4), and others more than 5 years (N=3). Many members had thus been onboarded at the end of the pandemic. The degree of familiarity differed across the team. Members within the sub-teams reported knowing each other well from before and being slightly less familiar with other sub-teams. Sub-team 3 had only been part of the team for 2 months when the data collection started. Two people had left the team in the last two years indicating a low turnover. When on-site, the teammates were located in an open office sharing with another team. The team comprised 3 sub-teams that historically dealt with similar tasks. Nine members had been pair programming before while others were either new to it or did not have developer tasks. Most of the team consisted of developers with either permanent positions (N=6) or hired-in consultants (N=9).

*Introducing PP*: To make sure everyone had the same understanding of the practice, two experienced developers working in-house (PP coaches) gave a short introduction to the whole team two weeks before the study started. The week the data collection started, the coaches and the researchers participated in the MC meeting, discussing with the team how to book pairing sessions, the importance of breaks when pairing, what tasks were suitable, the most typical barriers and enablers for PP, and how data would be collected by the researchers. During the four weeks of data collection (March 2023) the team members were encouraged to pair at least twice a week to pick up the practice. The coaches frequently interacted with the team during this period.

*B. Data collection and analysis*

To answer the research question, we collected both qualitative (interviews, observations) and quantitative data (surveys), see Fig.1. The quantitative data, the description of the variables, the interview guide, and the observation protocol are available in Appendix A. However, we cannot disclose the qualitative data (interview transcripts and field notes) as these materials may threaten the anonymity of the research participants.

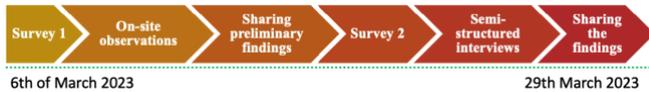

Fig. 1. Data collection procedures

*1) Qualitative data*

In terms of the qualitative data, we participated in and observed the introduction and tailoring of PP, conducted on-site observations of sessions, carried out 11 semi-structured interviews (Table I), and shared some of our findings with the team. We interviewed only 11 of 15 team members because we recruited the informants based on voluntary consent.

During the observations, we were focusing on how the teammates pair programmed and also fetching the team context (how the members were located, their surroundings and technical equipment, patterns of coordination and informal interaction, etc.). The interviews were conducted via video calls, with each interview lasting between 30 and 41 minutes. The interviews with developers focused on PP preferences in different work modes, the reasons for these preferences, and the informants' experiences with PP in general. The interviews with the team lead and the PP coach addressed the team context and their perception of the team and PP.

The authors used the observation and interview notes for the initial data analysis and for acquiring an overview of the data. Preliminary results were presented and discussed with the team and provided additional insight.

The interviews were transcribed resulting in 164 text pages. For this paper, all the sections of the transcripts related to PP in different work modes were thematically coded. Finally, we used Excel sheets to get an overview of the informants, their preferences for work mode in terms of PP, and their reasons for these preferences.

*2) Quantitative data*

The quantitative data was collected as part of a future bigger sample study that will include several teams. Therefore, reporting some of this data in the study does not pretend to achieve generalizability but rather indicates some trends among the participants that we used for structuring our interviews. The quantitative data was collected by administering a questionnaire in March 2023 when the data collection started (Survey 1) and after the four weeks of the team practicing PP (Survey 2). The responses were collected anonymously with each participant generating a unique ID. During both procedures, the respondents answered the questions on paper with researchers and the whole team present in the room. To ensure anonymity, the respondents were encouraged to place their questionnaires in a sealed envelope after completion.

TABLE I. OVERVIEW OF THE INTERVIEWED PARTICIPANTS

| ID | Employment[a] | # of days days off-site on a typical week | Work mode preference |
|---|---|---|---|
| **Dev1** | Permanent | Maximum 1 | On-site |
| **Dev2** | Consultant | 1 | Did not specify |
| **Dev4** | Permanent | 0 | On-site |
| **Dev5** | Consultant | 2 to 3 | No preference |
| **Dev7** | Consultant | 2 to 3 | No preference |
| **Dev8** | Permanent | 1 | Did not specify |
| **Dev9** | Consultant | 3 | Remote |
| **Dev11** | Consultant | 3 | On-site |
| **TeamL** | Consultant | 0 | N/A |
| **TechL** | Permanent | 1 to 2 | No preference |
| **Coach** | Permanent | N/A | N/A |

[a]Permanent in-house employee or hired-in consultants

The questionnaire items were concerned about how the PP was practiced and about teamwork. For the current paper, we only used the items related to practicing PP (Table II). The response rate was slightly higher in Survey 1 (100%) than in Survey 2 (93%). We asked about the number of the PP sessions and the length of sessions the previous week, the desired number of sessions, satisfaction with PP, and the intention to practice it. However, the data on PP was fragmented because not all the participants practiced PP, which resulted in some missing values. Additionally, we received only 9 responses on items (1) the number of sessions and (2) the length of sessions the previous week. Therefore, we carried out the statistical tests based solely on the available data (ignoring data from the informants whose responses were incomplete), which resulted in a lower N than the total of the team members (see Table II). We computed correlation effects (Pearson's r), paired t-tests to look for differences in practices and attitudes related to PP between Survey 1 and 2; and independent sample t-tests to explore group differences between consultants and in-house developers.

## IV. RESULTS

This section describes the results of our data analysis. We discuss the impact of different work modes on PP sessions and the various factors that influence team members' preferences, such as social connection, communication, workplace surroundings, and technical equipment.

*A. Hybrid work and PP sessions in the team*

The majority of team members (14 out of 15) were combining remote days with days at the office. Mondays and Tuesdays were the agreed-upon office days, while many were working remotely on the remaining days. On Thursdays, the majority of permanent employees were also present on-site to participate in the weekly competence development, whereas many consultants preferred working from home since they were not part of this event. On any given day, there were always some team members working on-site. In this way, the team was similar to what is described as a *partially aligned team* by Smite et al. [26], which is a form of hybrid work. In other words, the whole team was never fully on-site or fully remote on any given day.

The PP sessions were affected by hybrid work, so the members were alternating between the *on-site sessions* (both partners on-site), the *remote sessions* (both partners offsite), and the *mixed sessions* (one of the partners is on-site while the other one joins in remotely), see Table III. An average PP session length, regardless of the mode, was consistently reported to be about 2 hours (with the mean values ranging from 107 to 120 minutes across the measurements). However, the length varied, showing that some participants paired for about 30 minutes while others did for four hours and more. Discussing with the engineers, we learned that a reason for the variation could be people's availability. Those who had more meetings tended to have shorter PP-sessions while others could pair up for several hours in a row. Developer 9 described that he emerged in a state of flow when pairing: "*I think our tasks are a lot of fun, so suddenly we can end up coding for three hours, you don't even notice*".

TABLE II.  OVERVIEW OF THE PP IN THE TEAM AT THE TWO MEASUREMENT POINTS

| # | Item | Survey 1 | | Survey 2 | |
|---|---|---|---|---|---|
| | | N | M (SD) | N | M (SD) |
| 1 | Number of sessions last week (min.)[a] | 13 | 1.77 (1.42) | 9 | 4.22 (2.28) |
| 2 | Length of sessions | 13 | 117.69 (56.74) | 9 | 120.00 (80.78) |
| 3 | Desired number of sessions | 13 | 3.15 (1.09) | 13 | 3.50 (2.36) |
| 4 | Satisfaction with PP as practice | 13 | 4.38 (0.51) | 13 | 4.15 (0.55) |
| 5 | Intention to continue pairing in the future | 13 | 4.92 (0.28) | 13 | 4.85 (0.55) |

*Note. M – mean score, SD – standard deviation, N – number of valid cases.* [a]*Significant change (t (8) = -2.33, p = 0.05).*

As can be seen from Table II, the number of sessions per week increased from about 1.5 to 4 sessions, possibly as a result of the intervention that we witnessed. The interviews also indicated that many participants had more frequent sessions by the end of the data collection. Developer 7 commented on the effect of managers promoting PP: "*We did it [PP] also before but not as much. So now we got a push to do it*».

When being on-site, we observed 8 PP sessions in two days (Fig. 2). In line with the quantitative findings, the observed sessions lasted 1.5 hours on average (ranging between 30 minutes and 4 hours). The team members were using the PP sessions to solve a variety of tasks that were typical for this platform team, such as bug fixing, application migration, updating, and testing. During a session, the driver (the one operating at the keyboard) would typically have less competence in solving the task at hand than the navigator (the guidance and feedback provider), as had been recommended by the PP coaches. In a pair where the role distribution was different, the partners would swap roles whenever needed (e.g., the driver being tired), otherwise the roles remained unchanged throughout a session. The teammates were also encouraged to frequently change partners, nevertheless, we observed that many developers preferred pairing with one or two partners that they knew well. As developer 8 commented in the subsequent interview: "I must admit I prefer [to code with] a few people, with those I know. […] And I notice this tendency in the whole team". The time of the on-site sessions was roughly distributed between explaining and decision-making (about 80%), waiting for the test response (10%), and code writing/editing (10%). The interviews and the feedback sessions suggested that this had to do with the team tasks (e.g., application migration) that required a deep understanding of how to solve the problem prior to actually solving it.

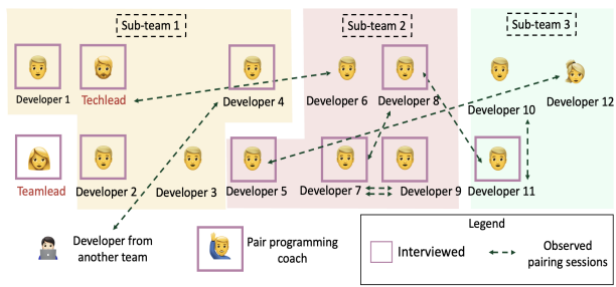

Fig. 2 Observed pair programming sessions on-site

We found that the team members had different work mode preferences depending on whether they were hired-in consultants or permanent employees. The group comparison showed that the consultants on average reported working remotely slightly more often (2.56 days during a typical week) than the permanent employees (about 1.4 days). The interviews revealed that a reason was the weekly competence day for the permanent employees.

Interestingly, we found that those who worked from home reported more PP sessions. We registered a correlation between the number of days at the office and the number of PP sessions per week (Survey 1), $r = 0.57$, $p < 0.04$. Although we did not find such correlation in the subsequent data collection (Survey 2), we decided to follow up on this topic in the interviews. The responses revealed that a proportion of the interviewed developers indeed had preferences in terms of the work mode when it comes to pairing (see Table I). One developer had a strong preference towards pairing from home: "*PP works great at home, I think it is much easier than at the office*" (Dev9). Three informants preferred on-site, while the remaining informants did not have a specific preference or did not indicate it. Two of them (TechL, Dev5) explained that which work mode was beneficial for PP depended on the nature of the task, for example, PP was preferred if the task was complicated.

B. *Benefits and challenges of PP sessions in different work modes*

Since the informants preferred some work modes over others, we explored the reasons for these preferences, which resulted in a list of challenges and benefits (summarized in Table III).

**The social connection.** The opportunity to stay connected with peers was reported to be a benefit both on-site and remotely (see Table III). One informant explained: *"It is nice to be at home and pair program. One may chat a bit. You get the office feeling, also with small talk, which I think is good"*. (Dev9). Interestingly, another interviewee highlighted the social aspect as a reason for PP on-site and explained that the reason for preferring PP in the office was that it was *"nicer and more social"* (Dev11).

**Communication.** Communication during the PP session was also affected by the session's mode. One participant reported that pairing remotely was challenging for effective problem-solving as the pair often ended up working in parallel: "*It's chaotic when you pair program remotely […]* we both begin digging in our own screens and then we don't communicate much" (Dev1). Another emphasized the benefits of using a physical whiteboard to facilitate discussions as an advantage of PP in the office. For him, drawing helped communicate when discussing complicated problems: "*The jobs that don't require much discussion and explanation can be hybrid. But if one needs to draw something, it is better on a [physical] board*" (TechL).

TABLE III. BENEFITS AND CHALLENGES OF THE PP-SESSIONS IN DIFFERENT WORK MODES

| Mode | Benefits | Challenges |
|---|---|---|
| **Remote** (Both partners join remotely) | Staying connected with peers, Ability to solve multiple problems at the same time, Ability to edit code simultaneously, Non-intrusive for other colleagues | Reduced communication during problem-solving |
| **On-site** (Both partners join on-site) | Staying connected with peers, Ability to use the whiteboard | Interrupting nearby colleagues, Reduced availability of meeting rooms, Meeting rooms not fully suitable for PP |
| **Mixed** (One partner on-site, the other remote) | | Interrupting nearby colleagues (when using microphone), Having to work from the meeting room (when one partner is on-site) |

**Workplace surroundings.** Various aspects of the workplace surroundings seemed to be relevant for PP in the office. Three informants acknowledged fearing interrupting other colleagues in the open office. Notably, it was especially a problem for the hybrid sessions when one of the partners was joining from home: "*If one is at the office, then one goes to a quiet room if one needs to talk to someone who is not there. I feel PP works best home-home or work-work*" (Dev7). Oddly, we did not acquire direct evidence of other members being bothered by the PP sessions nearby. On the contrary, two interviewees who paired in the open office (Dev2, Dev8) explicitly admitted that the sound level during the teammates' sessions was not problematic for them. However, several indirect data points suggested that the sound level may still be challenging in some settings. During our presentation of the preliminary findings for the team, some stated that several PP sessions nearby do create noise and that sometimes one feels "dragged into" the discussions. This issue may also be related to the office environment. During the onsite visits, we observed that the meeting rooms were not always available nearby, which was later described as a potential problem by Dev2. This indicates that the success of PP in the open office may either depend on personal preferences or on the office layout.

**Technical equipment**. The workplace setup also influenced the informant's impression of the PP. One informant explained that when working remotely, each one has their own screen, which made work more effective: "*Even though you share a screen [off-site], your partner can still do some searching on his own, so it becomes rather effective. In the office, you have to come back to your desk to do the search*" (Dev7). Another added that remote PP enables simultaneous code editing: "*Then each one has their own equipment and can work on the parallel tasks*". (Dev9) Interestingly, working in parallel was described as a challenge for problem-solving by Dev1, which seems contradictory. Lacking equipment in meeting rooms also seemed to be problematic for some, as we learned during the feedback session (e.g., due to only a single screen).

## V. DISCUSSION

PP is one of the collaborative practices that are known for numerous benefits, given the sessions are performed according to recommendations [21]–[23]. Although we now have an increasing knowledge of remote PP (including that during the COVID-19 outbreak), studies of the practice during hybrid work are less common. Therefore, we in this study presented preliminary results to answer the RQ: *How is PP practiced in hybrid work*? We found that during hybrid work, the PP sessions in a team can occur in three different work modes: on-site, remote, and mixed. We will first discuss our findings in all three of these modes and then proceed to practical implications for organizing the hybrid work for successful PP.

**On-site PP-sessions.** We observed many on-site PP sessions in the team and found that when the partners were pairing in the office, they used 80% of the time for explanations and decision-making and only about 10-20% for code editing. These findings seem to be due to the specifics of the studied team (the platform team) that was responsible for the migration of the applications to the cloud. These tasks were described as not straightforward and thus requiring additional discussions on how to proceed. Although this may not seem like a typical scenario for PP, our results are also in line with the findings of Zieris and Prechelt [21], who showed that explaining and decision-making were central dialog modes for PP.

Another interesting finding is that the PP partners often relied on distinct roles (the navigator and the driver) when there was a competence gap between the partners, which is classical for PP [18]. In contrast, when both partners had the same competence level they had a more flexible role distribution, as found by Chong and Hurlbutt [19], and then had a more fluent way of working. We, therefore, conclude that the competence level might affect how pairs distribute their roles during a PP session.

Finally, our results indicate that pairing in the on-site mode has both benefits and challenges. Although the informants regarded PP in the office as positive due to the possibility of staying connected with peers and using physical artifacts, several drawbacks related to the open office (e.g., noise level and the fear of disturbing others) were also reported. Earlier research shows that open offices reduce face-to-face communication between employees [33]. At the same time, an open office might be beneficial for teams because the members can spontaneously be involved in PP sessions, as indicated by one of our informants.

**Remote PP sessions.** Based on the interviews and the survey data we observed that several couples paired from home and that one informant even preferred remote PP sessions to on-site ones. One rationale for this was the increased efficiency due to the ability to solve parallel tasks and edit code simultaneously. Interestingly, the same rationale was described as problematic by others because it impaired collaboration leading to the pair working independently. This discrepancy may be due to personal preferences. However, working independently during a session can lead to negative outcomes (e.g., impaired knowledge transfer [21] due to reduced communication) and should thus be applied with caution.

Furthermore, the participants indicated clear preferences in terms of PP partners preferring people they knew well from before. Being familiar with the partner and having established social connections from before are among the prerequisites of successful remote PP [5], [29]. On the other hand, sticking to the same partners may reduce knowledge transfer [21] in a team due to reduced exchange between the members. At the same time, communication and psychological safety can decrease during remote meetings [3]. Therefore, pairing with a few well-known partners may be an adaptive strategy when it comes to remote PP sessions.

**Mixed PP sessions** (when one partner joins from the office, the other from elsewhere) are likely to occur in the context of hybrid work because work location and/or hours of individual employees are not strictly standardized [4]. The team members in our study were in the office for 2-3 days a week while working remotely on the remaining days, which is similar to other findings on hybrid work [24]. Our findings indicate that mixed sessions may be the least advantageous of all three types. The major drawback was related to the on-site partners being urged to move to a meeting room when being contacted by the remote counterpart. Whereas both on-site or fully remote PP sessions could be positively experienced by the participants, no advantages were described for the mixed sessions. This least advantageous evaluation of mixed PP sessions is similar to the findings of Tkalich et al. [3], who found that the mixed mode can be problematic for software teams because it may lead to the alienation of the remote members. However, our findings indicate that mixed PP sessions might be problematic only for the teams working in open offices. In the future, we plan to investigate whether it is also the case for teams in other office settings.

**How should PP sessions be organized in the context of hybrid work?** Our findings have important implications for companies operating in the context of remote work. The results reveal that the hybrid work itself does not necessarily threaten the success of PP but that several aspects are to be treated with caution. Firstly, we found that PP sessions in all work modes (on-site, remote, and mixed) may be challenging for various reasons. Even pairing on-site may disturb other colleagues or can be uncomfortable due to sub-optimally equipped meeting rooms. This highlights that successful PP requires preparation and suitable infrastructure. One recommendation for management can be to make sure that the office space supports the practice and that several members can PP during the same period. Further, both on-site and remote modes of PP may strengthen teamwork. The participants reported feeling more connected with peers regardless of whether the PP happened on-site or remotely. Earlier findings show that during hybrid work teams may experience poorer team cohesion, coordination, and psychological safety [3], [25]. Therefore, PP can be used as a team-building intervention, which is especially valuable in the context of hybrid work. Considering how to utilize PP to nurture the social dimension in both on-site and remote settings could be valuable, particularly for companies with employees in multiple physical locations.

The finding that the remote PP sessions can be equally successful as the on-site sessions and that some developers prefer the remote mode is surprising given that earlier results indicated the opposite [5], [15]. The discrepancy can be explained either by the difference in contexts ([5] investigation took place during the COVID-pandemic) or by the individual preferences of the developers. Indeed, the team members had varying preferences for the work mode. This may serve as a warning for employers to make sure different needs are satisfied if they wish their employees to succeed with PP. For example, meeting rooms or cell offices are necessary for comfortable sessions on-site if the practice is to be scaled for numerous employees in a team. Perhaps some guidelines should be established to avoid sub-groups as we know that this can be a risk during hybrid work [3]. As workers post-pandemic desire flexibility in their work situation, PP in the mixed work mode is likely to become even more widespread. This may, however, be a challenge, as suggested by our findings. To ensure good conditions for PP, we recommend teams and their managers explore the possibilities for aligning their office presence to facilitate either purely on-site or remote PP sessions.

*A. Limitations and long-term plan*

An apparent limitation of this emerging results study is that it is based on data from a single team which limits the generalizability of the findings. In the future, we plan to collect similar data from additional teams. Second, some participants were novices in PP which may influence the way they practiced it. Therefore we will repeat the study in teams more experienced in PP. Third, the quantitative data is limited and should thus be treated as an additional insight into other data sources. Nevertheless, it shows some trends, which can be explored in further studies. Finally, the observations were conducted when most of the team was co-located, therefore we did not observe remote and mixed sessions – we are planning to do this as a next step. To summarize, our future research objectives include 1) collecting data on PP in various modes from additional teams (estimated by the end of 2024) and 2) carrying out additional data collection focusing on remote and mixed PP sessions (end of 2024).

VI. CONCLUSIONS

Companies worldwide are now adapting to hybrid work, and management and teams ought to consider conditions that foster well-functioning PP practices in these new settings. Our preliminary results indicate that PP can be practiced through on-site, remote, and mixed sessions, where the mixed mode is the least advantageous. The teams intending to PP are thus recommended to align their office presence to avoid mixed PP sessions. The findings also highlight the importance of adapting the work environment to suit onsite just as remote PP sessions. Generally, this study is a first step in evaluating how the PP practice can be introduced in hybrid settings. We hope that our study design can inspire other researchers to explore and evaluate alternatives for implementing PP. In the future, we will build on these preliminary findings to explore how PP can be introduced in other teams and organizations.

**Appendix A.** Available at
https://doi.org/10.5281/zenodo.8087197